\documentclass[runningheads]{llncs}
\usepackage{graphicx}
\usepackage{placeins}
\usepackage{float} 
\usepackage[utf8]{inputenc}
\usepackage[T1]{fontenc}
\usepackage[portuguese]{babel}
\usepackage{icomma} 
\usepackage{amsmath,amssymb,amsfonts} 
\usepackage{listings}
\usepackage[portuguese, ruled]{algorithm2e}
\usepackage{xpatch}

\begin{document}
\title{Modelagem de Sistemas Audiométricos Usando Técnicas de Computação Flexível}
%
%
\author{Erick Schultz S. A. Caetano\inst{1},
Denise Fonseca Resende\inst{2},
Samir Angelo Milani Martins\inst{1,2},
Erivelton Geraldo Nepomuceno\inst{1,2} e
Leonardo Bonato Felix\inst{1,3}
}
\authorrunning{E. Schultz et al.}
%
\institute{Programa de Pós-Graduação em Engenharia Elétrica (UFSJ/CEFET) \\ 	Departamento de Engenharia Elétrica UFSJ - Universidade Federal de São João del-Rei. São João del-Rei, MG, Brasil \\
\email{schultz.s.a.c@gmail.com}\\
\and
Grupo de Controle e Modelagem, Departamento de Engenharia Elétrica, Universidade Federal de São João Del-Rei, São João del-Rei, MG, Brasil\\
\email{denisesade@gmail.com,martins@ufsj.edu.br,nepomuceno@ufsj.edu.br}
\and
Departamento de Engenharia Elétrica, Universidade Federal de Viçosa (UFV), Viçosa, MG, Brasil
\\
\email{leobonato@ufv.br}
}
\maketitle              

\begin{abstract}
 Em identificação de sistemas os fenômenos estudados vêm acompanhados de incertezas, sejam elas decorrentes dos dados de medição ou cálculos computacionais. Os dados com valor intervalar oferecem uma maneira valiosa de representar as informações disponíveis em problemas complexos nos quais a incerteza, a imprecisão ou a variabilidade devem ser levadas em conta. O presente trabalho visa determinar parâmetros intervalares para um modelo considerando incertezas de medição nos dados utilizando o estimador MQ e redes neurais. O principal objetivo do trabalho é aplicar esta técnica em sistemas audiométricos, particularmente na detecção automática de respostas auditivas levando em consideração os conceitos de computação flexível que oferecem soluções tolerantes à subjetividade, imprecisão ou incerteza. 		 
Foi possível obter um modelo com parâmetros intervalares que permite um conjunto infinito de parâmetros ser avaliado como um intervalo limitado.
O modelo foi validado utilizando dois métodos, um com predição de 2 passos à frente, utilizando o estimador MQ e estendido para outra representação como redes neurais 5 passos de atraso. Foi possível demonstrar que os parâmetros intervalares geram resultados também intervalares que contém a maioria dos dados de validação aprimorando a confiabilidade do sistema.
\keywords{Computação Flexível, Aritmética Intervalar, Identificação de Sistemas, Sistemas Audiométricos.}
\end{abstract}
\section{Introdução}
O uso de modelos não lineares em identificação e análise de sistemas, em muitos casos são essenciais pois, a maioria dos fenômenos observados na prática surgem da não linearidade do sistema original \cite{AguirreLuisAandBillings1995}. Em geral, processos não lineares podem ser caracterizados em identificação de modelos não lineares com incertezas \cite{bil80}. A identificação de sistemas tem sua relevância, pois muitos processos tem características não estacionárias, incluindo um grande número de sistemas físicos, fisiológicos e bioquímicos \cite{He2013}.

Muitos sistemas estão sujeitos a erros decorrentes da aquisição dos dados ou até mesmo de arredondamentos feitos pelo computador \cite{He2013,Nepomuceno2016a,IEEE2008}. Com o intuito de obter resultados mais confiáveis, a análise intervalar juntamente com a computação flexível (\textit{soft-computing}) tem se tornado campo ativo de pesquisa e aplicação \cite{Lopez-Guede2019,Leal,Com2013}.

Muitas áreas de aplicação em tempo real relacionadas com \textit{soft-computing} são diagnóstico de falhas, análise de dados, otimização, controle, \textit{data mining}, reconhecimento de padrões, processamento de imagens, processamento de sinais, sistemas de tráfego e transporte, estimativa de parâmetros, solução robusta, sistema adaptativo, identificação de sistemas, auto-organização, otimização multiobjetivo, análise de falhas, etc \cite{Iftikhar2018,Kumari2018}.
	
Tendo em vista a combinação da análise intervalar e identificação de sistemas com métodos de \textit{soft-computing}, essas ferramentas mostram seu potencial em sistemas audiométricos e em redes neurais por exemplo, satisfazendo os requisitos dos sistemas ao lidar com situações complexas, das quais exigem um grau de precisão maior \cite{HuChenyi2008,Roque2007}. A coleta ou geração de dados pode trazer agregadas a elas incertezas decorrentes dos dados de medição ou cálculos computacionais que, podem subjetivar os resultados levando a uma informação incompleta, parcial ou conflitante da verdade \cite{Pedrycz2001,Shao2013}.

A abordagem neste artigo consiste na estimação dos parâmetros intervalares utilizando o estimador de mínimos quadrados (MQ) para os modelos polinomiais \cite{Aguirre2007,Nepomuceno2007a} e estendido para redes neurais artificiais \cite{Martins2013,Zhang2016}. Assim, será possível empregar o \textit{software} MATLAB, em que pretende-se analisar os dados, obter o modelo e através da \textit{toolbox Intlab} \cite{Hargreaves2002,Rump1999} considerar valores medidos como intervalos e assim obter os parâmetros do modelo também intervalares.
	
O restante do artigo está organizado da seguinte forma. Os conceitos preliminares são descritos na Seção 2. A Seção 3 apresenta a metodologia do presente trabalho. A Seção 4 apresenta os resultados obtidos. A conclusão é apresentada na Seção 5.
	
\section{Conceitos Preliminares}
	
\subsection{Modelos NAR Polinomial}
	
Um modelo NAR (auto regressivo não-linear) pode ser representado como (\ref{eq:15}) \cite{LEONTARITIS1985}.
\begin{equation}
\label{eq:15}
y(k)= F^l[y(k - 1), ..., y(k-ny)]+e(k)
\end{equation}
em que $y(k)$ representa o sinal de entrada, $e(k)$ representa o ruído e erros associados ao sistema e $F[\cdot]$ é uma função polinomial não linear de grau $l$.
	
\subsection{Critério de Informação de Akaike}
	
A identificação de modelos é uma aproximação da realidade, pois existe perda de informação. Para explicar o estudo em questão deve-se selecionar o modelo com melhor desempenho entre os que foram ajustados. Para isso critérios de informação são utilizados para estimar a ordem ou o número de termos de um modelo. 
Akaike (1974) propôs o \textit{critério de informação} que estima o número de regressores que devem ser incluídos no modelo, sendo definido por:
	\begin{equation}
	\label{eq:5}
	AIC(n_\theta) = Nln[\sigma_{erro}^2(n_\theta)] + 2n_\theta,
	\end{equation}
	em que $N$ é o número de dados, $\sigma_{erro}^2(n_\theta)$ é a variância do erro de modelagem e $n_\theta = dim[\hat{\theta}]$ é o número de parâmetros no modelo.
	
A Equação (\ref{eq:5}) explica que a medida que termos são incluídos o número de graus de liberdade aumenta, ou seja, a medida que $n_\theta$ aumenta, $\sigma_{erro}^2(n_\theta)$ diminui. Portanto, se o ``custo'' de incluir um determinado termo em $2n_\theta$ exceder a redução em $ln[\sigma_{erro}^2(n_\theta)]$ o termo não deve ser incluído no modelo.
	
\subsection{Aritmética Intervalar}
	
Baseado em uma extensão do sistema dos números reais, Moore et al. (1958) foi um dos pioneiros na área da análise intervalar. 	O propósito de Moore foi colocar limites em erros de arredondamentos nos cálculos, onde assim poderia considerar resultados mais confiáveis.
	
Considerando então um intervalo sendo um subconjunto não-vazio dos números reais  $\mathbb{R}$, o intervalo é definido como: 
	\begin{equation}
	[x] \equiv [\underline{x},\overline{x}] := \{{x \in \mathbb{R}| \underline{x} \le x \le \overline{x}}\},
	\end{equation}
em que $\underline{x}$ é o limite inferior e $\overline{x}$ é o limite superior do intervalo $[x]$.
	
As operações básicas aritméticas quando processadas em aritmética intervalar fornecem intervalos de modo a conter todos os valores possíveis. Sendo assim, dado $\text{x} = [\underline{x}, \overline{x}]$ e $\text{y} = [\underline{y}, \overline{y}]$, as quatro operações elementares são definidos por 
\begin{equation} 
	\label{eq:4}
	\begin{array}{cccc} 
	x + y = [\underline{x} + \underline{y}, \overline{x} + \overline{y}]\\
	
	x - y= [\underline{x} - \underline{y}, \overline{x} - \overline{y}]\\ 
	
	x \times y = [min(C), max(C)]\\ 
	
	1/x = [1/\overline{x}, 1/\underline{x}] \,\, se \,\, \underline{x} > 0 \,\, ou \,\, \overline{x} < 0.
	\end{array} 
\end{equation}
\noindent em que $C$ é o conjunto $\{\underline{x} \underline{y},\underline{x} \overline{y}, \overline{x}\underline{y}, \overline{x} \overline{y}\} $. O problema das operações básicas intervalares (\ref{eq:4}) ocorre quando aplicadas no computador, o mesmo não respeita os axiomas definidos. Em consequência deste problema propriedades básicas da aritmética dos números reais como: associativa, comutativa, inversas e distributivas não são garantidas para a aritmética de ponto flutuante \cite{IEEE2008}.

\subsection{Computação flexível}

\textit{Soft computing} mudou a forma do mundo profissional fornecendo resultados aproximados da verdade e esta diretamente ligada a aritmética intervalar. Este método é tolerante à imprecisão, aproximação e incerteza. É baseado em redes neurais, lógica difusa e raciocínio probabilístico \cite{Society2015}. 

A computação flexível aprende com dados anteriores e prevê o futuro. A computação convencional fornece uma solução usando uma solução matemática, mas a computação flexível fornece uma solução aproximada da realidade. \textit{Soft computing} é usado em sistemas evolutivos, redes neurais, e muitos outros problemas complexos podem ser resolvidos, sendo estes, não solucionáveis usando métodos analíticos convencionais.

Computação flexível é uma combinação de lógica difusa, algoritmos genéticos, redes neurais artificiais, aprendizado de máquina e raciocínio probabilístico. Algumas vantagens da utilização deste método são: Podem gerar seus próprios programas; Pode lidar com dados ambíguos e ruidosos; Permite cálculos paralelos; Os programas aprendem por conta própria; As aplicações são tolerantes à imprecisão, incerteza e aproximação; Requer muito menos tempo para simulação; Seu modelo é o cérebro humano.

Como o computador apresenta limitações físicas, ou seja, problemas de armazenamento o conjunto de números que ele representará será finito.
A utilização das técnicas de \textit{Soft computing}, que é uma coleção de metodologias que visam explorar a tolerância para imprecisão e incerteza para alcançar a tratabilidade, robustez e baixo custo da solução, podem auxiliar nos problemas de arredondamento nos computadores e a utilização da representação por intervalos nas operações pode melhorar os resultados, principalmente para números que ultrapassam o limite de precisão da máquina.

\subsection{Estimador de Mínimos Quadrados}
	
O estimador de mínimos quadrados é utilizado para a determinação dos parâmetros do modelo estimado \cite{Aguirre2007}.	
O método de estimação por mínimos quadrados consiste em minimizar o quadrado das diferenças entre os valores observados de uma amostra e seus respectivos valores esperados.	
Portanto, uma massa de dados pode ser representada por uma Equação matricial da forma
	\begin{equation}
	\label{eq:6}
	y = \Psi \hat{\theta} + \xi
	\end{equation}
\noindent em que $\Psi$ é chamada matriz de regressores e $\xi$ é o resíduo.
	
Além disso, o estimador MQ pode ser representado por
\begin{equation}
	\label{eq:7}
	\hat{\theta}_{MQ} = [\Psi^T \Psi]^{-1} \Psi^T y
\end{equation}
em que $\hat{\theta}_{MQ}$ são os parâmetros estimados por mínimos quadrados.

\subsection{Determinação da Estrutura e Validação}
	
Para resolver o problema de determinação de parâmetros intervalares optou-se primeiramente por realizar a identificação de um modelo e posteriormente adicionar um intervalo para a saída, desse modo evita-se o crescente erro computacional associado a cada operação matemática. 
	
O conjunto de termos candidatos foi escolhido considerando-se grau de não linearidade 4 
e máximo atraso de saída 4. Posteriormente calcula-se a taxa de redução de erro (ERR) para todos estes termos de acordo com a Equação (\ref{eq:err}) \cite{Korenberg1988,Billings1989} .
	\begin{equation}
	\label{eq:err}
	ERR_i = \frac{g_i^2 \langle \Omega_i,\Omega_i \rangle }{\langle y,y \rangle }
	\end{equation}
em que $\langle \cdot\rangle $ é o produto interno, $\Omega_i$ é o i-ésimo regressor ortogonal e $g_i$ o seu respectivo parâmetro.
	
Em seguida utiliza-se um algoritmo de determinação de estrutura baseado no AIC. Nesse algoritmo primeiramente determina-se todos os parâmetros dos termos candidatos através de um estimador de mínimos quadrados ortogonais e em seguida utiliza-se para compor o modelo o número de termos selecionados pelo AIC. 
	
Para validar o modelo utiliza-se a predição 2-passos à frente afim de verificar a capacidade de adequação do modelo ao conjunto de dados de validação. O desempenho do modelo é quantizado pelo índice RMSE (erro quadrático médio normalizado) descrito em (\ref{eq:rmse}).
\begin{equation} \label{eq:rmse}
RMSE= \frac{\sqrt{\sum_{k=1}^{N}(y(k)-\hat{y}(k))^{2}}}{\sqrt{\sum_{k=1}^{N}(y(k)-\bar{y})^{2}}}
\end{equation}
em que $\hat{y}(k)$ é a saida do modelo e $\bar{y}$ é o valor médio da saída $y(k)$ calculado sobre o conjunto de dados de identificação. A análise de resíduos é calculada a partir de ($\xi = y - \Psi \hat{\theta}$), sendo utilizada para verificar se os parâmetros do modelo identificado foram ou não estimados corretamente. Essa análise indica se o modelo foi capaz de explicar tudo que era possível no conjunto de dados, para isso utiliza-se as Equações (\ref{eq:10}) $-$ (\ref{eq:12}), \cite{Aguirre2007}.
\begin{equation}
	\label{eq:10}
\scriptstyle 	r_{\xi \xi}(\tau) = E[\xi(k) - \overline{\xi(k)}(\xi(k - \tau) - \overline{\xi(k}))]= \delta (\tau),
\end{equation}
\begin{equation}
	\label{eq:11}
\scriptstyle 	r_{\xi \xi^{2'}}(\tau) = E[\xi(k) - \overline{\xi (k)}(\xi(k - \tau) - \overline{\xi^{2(k)}})]=0,
\end{equation}
\begin{equation}
	\label{eq:12}
\scriptstyle 	r_{\xi^{2'} \xi^{2'}}(\tau) = E[\xi^{2}(k) - \overline{\xi^{2}(k)}(\xi^{2}(k - \tau) - \overline{\xi^{2(k)}})]=\delta (\tau).
\end{equation}

\section{Metodologia}
	
\subsection{Definição do Problema}
	
Neste artigo pretende-se identificar os parâmetros do sistema na forma intervalar, de tal forma a incorporar incertezas relativas aos erros de medição dos dados. Afim de alcançar este objetivo a primeira etapa é identificar um modelo para o sistema sem considerar as incertezas numéricas de medição. 
Os dados a serem identificados foram capitados do Eletroencefalograma (EEG) de 10 voluntários, sendo 6 do gênero feminino e 4 do masculino, foram coletadas em uma cabine acusticamente isolada. Os dados utilizados foram adquiridos do trabalho \cite{Silva2016a}, em que, a coleta dos mesmos não é o enfoque deste trabalho.
	
Para a aquisição dos sinais utilizou-se um amplificador de sinais biológicos de 36 canais configuráveis, modelo BrainNet BNT 36 (fabricado pela empresa Lynx Tecnologia), compatível com eletroencefalografia  \cite{Silva2016a}.
O eletroencefalógrafo é composto por 22 canais monopolares com referência comum, 10 configuráveis mono ou bipolares e 4 entradas RCA configuráveis como entrada ou saída de dados. 
	
Nos experimentos realizados utilizou-se os 22 canais monopolares, além do terra e uma entrada RCA (Radio Corporation of America) como o trigger.
Foram realizadas 48 sessões de coleta por voluntário (seis intensidades para cada um dos quatro tons por orelha). Cada sessão foi composta de 352 janelas, cada uma com 1.024 pontos, de 9 minutos e 58 segundos de estimulação monaural, realizadas nas intensidades de 15, 20, 25, 30, 40 e 50 dB SPL para as frequências portadoras de 0,5, 1, 2 e 4 kHz, respectivamente. Portanto, cada voluntário se submeteu a um exame de 7 horas, 58 minutos e 24 segundos.
	
\begin{figure}[htb]
	\begin{center}
		\includegraphics [scale=0.45]{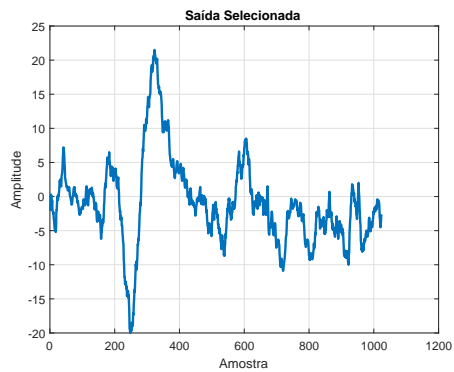}
		\caption{Saída do eletrodo número 9 do paciente escolhido com intensidade de 15 dB, da orelha direita de um único paciente. 
		}
		\label{saida}
	\end{center}
\end{figure}

Neste trabalho foi selecionado a intensidade de 15 dB, da orelha direita de um único paciente.
Além disso, as janelas não foram utilizadas para o estudo em questão. 
Para realizar a identificação optou-se por usar uma única saída e nenhuma entrada, ou seja, um modelo NAR. A saída escolhida se encontra na Figura \ref{saida}.

É importante verificar se o sinal está superamostrado e caso necessário fazer a decimação do mesmo utilizando o seguinte critério $10\Delta \leq \tau _{m}\leq 20\Delta$, onde $\tau _{m}$ é o valor de trabalho e $\Delta$ é a taxa de decimação. Esse valor de trabalho é obtido da análise das Equações (\ref{eq:8}) e (\ref{eq:9}), onde $\tau _{m}$ é o primeiro mínimo das funções de autocovariância linear e não-linear da saída \cite{Aguirre2007}. O conjunto de dados obtido foi dividido entre dados de identificação e dados de validação, sendo 50$\%$ para cada.
\begin{equation}
\label{eq:8}
\scriptstyle r_{y^*}(\tau) = E[y^*(k) - \overline{y^*(k)}(y^*(k - \tau) - \overline{y^*(k}))]
\end{equation}
\begin{equation}
\label{eq:9}
\scriptstyle r_{y^{*2'}}(\tau) = E[y^{*2}(k) - \overline{y^{*2}(k)}(y^{*2}(k - \tau) - \overline{y^{	*2(k)}})]
\end{equation}

\subsection{Parâmetros Intervalares}
	
Utilizando a \textit{toolbox Intlab} do \textit{software Matlab} é possível determinar valores intervalares. Esta ferramenta é utilizada para a aritmética intervalar, pois suporta intervalos reais e complexos, escalares, vetores e matrizes. Neste trabalho o comando \textit{midrad} do \textit{Intlab} é utilizado para a geração do intervalo, em que ``mid''  é o valor médio e ``rad'' é o raio.
	
Levando em consideração a sensibilidade do eletroencefalógrafo que é de ($1 \mu V$), este mesmo valor foi empregado como raio do modelo. A sensibilidade é a distância que o eletroencefalógrafo esta dos seus valores nominais, ou seja, é a altura do sinal em relação à sua amplitude, este teste é realizado na calibração do aparelho. Portanto, o comando \textit{midrad} é aplicado da seguinte forma:
\begin{equation}
\label{eq:16}
midrad(saida9d ; 0,1\mu)
\end{equation}
em que ``$saida9d$'' é a saída decimada do eletrodo número 9, representada na Figura \ref{saida} e ``$0,1$'' é a sensibilidade do eletroencefalógrafo, respectivamente o valor médio e o raio.
Como consequência desta aplicação, todos os posteriores valores serão intervalos, com raio de ($1 \mu V$), assim como a saída do sistema.
	
Para a estimação dos parâmetros intervalares utilizou-se o estimador de mínimos quadrados, onde para o cálculo da $\Psi$ utilizou-se os valores intervalares da saída. Aplicando a $\Psi$ intervalar na Equação  (\ref{eq:7}) obtêm-se um $\hat{\theta}_{MQ}$ intervalar estimado, do qual deve conter o valor verdadeiro do parâmetro.	

\subsection{Modelo NAR por redes neurais}

Outro método de modelagem aplicado, para fins de comparação com o modelo intervalar, foi um  modelo NAR por redes neurais. Foi utilizada a \textit{toolbox nntool} de redes neurais do \textit{software Matlab} com 5 passos de atraso.

A função \textit{trainlm} foi utilizada com o intuito de determinar as funções de treinamento e de desempenho da rede. A função com melhor taxa de convergência para casos de representação de relações entrada-saída é escolhida como candidata e o erro RMSE é calculado para avaliar a dispersão entre os dois modelos.

Uma rotina que avalia os resultados dos modelos gerados, variando os neurônios da camada escondida entre 10 e 30 neurônios, retorna o melhor resultado. Nesses testes os dados foram divididos em 60\% para validação e os demais foram divididos igualmente entre teste e validação. 
	
\subsection{Rotinas}
	
Todas as rotinas são executadas no Matlab 9.0 (R2010a). Foi utilizado um computador com um processador Intel Core i5-5200U CPU @ 2.20GHz e um sistema operacional \textit{Windows 10 Home} 64 bits. O Algoritmo (1) mostra o caminho traçado para a obtenção dos resultados.

\begin{algorithm}[H]
\label{Algoritmo_1}
{\textbf{Entrada:} Carrega dados}\\
{\textbf{Saída:} Modelo Intervalar}\\
1. Separa dados de identificação e dados de validação; \\
2. {\textbf{Função:}} auto covariância \% Verifica a auto covariância da saída \\
3. Decimação dos dados; \\
4. {\textbf{Função:}} gera termos \% Gera os termos candidatos \\
5. {\textbf{Função:}} akaike \% Critério de Akaike \\
6. Calcula a taxa de redução de erro para todos os termos; \\
7. {\textbf{Função:}} predição \% Predição k passos à frente \\
8. Análise de Resíduos; \\
9. Cálculo do RMSE; \\
10. Analise intervalar; \\
11. Método de mínimos quadrados; \\
12. Predição k passos à frente; \\
13. Método das redes neurais; \\
14. Cálculo do RMSE intervalar;
\caption{Modelagem de Sistemas Audiométricos}
\end{algorithm}

\section{Resultados}
	
\begin{figure}[htb]
	\begin{center}
		\includegraphics [scale=0.5]{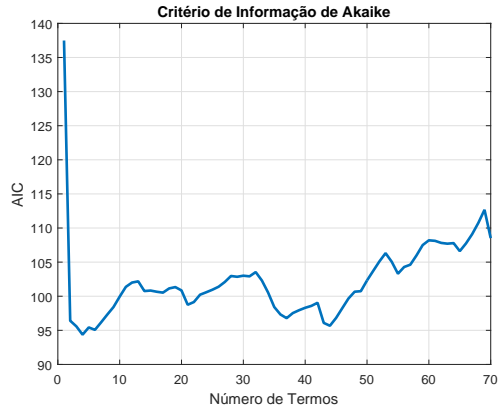}
		\caption{Critério de informação de Akaike (AIC) calculado mediante a inserção dos termos, classificados por meio da taxa de redução de erro (ERR). O número ótimo de acordo com o AIC foi de 4 termos, que são os 4 primeiros pontos da curva. }
		\label{aic}
	\end{center}
\end{figure}

No presente trabalho foi realizada a identificação do sistema considerado, utilizando um modelo NAR obtido aplicando ERR em que, a Tabela (\ref{tab:tabela1}) mostra os valores de ERR para os 10 primeiros regressores, e o critério de Akaike, do qual foram obtidos 4 termos para o modelo, como mostra a Figura \ref{aic}. 

\begin{figure}[H]
	\begin{center}
		\includegraphics [scale=0.5]{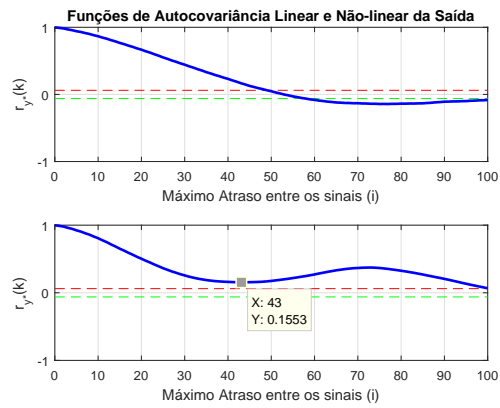}
		\caption{Autocovariância da Saída. Como o sinal estava superamostrado foi necessário realizar a decimação do mesmo. O critério utilizado foi $10\Delta \leq \tau _{m}\leq 20\Delta$, onde $\tau _{m}$ é o valor de trabalho e $\Delta$ é a taxa de decimação. O valor de trabalho  $\tau _{m}$ é o primeiro mínimo das funções de autocovariância linear e não-linear da saída. Pela figura este valor é de 43 amostras.
		}
		\label{autocov}
	\end{center}
\end{figure}

Primeiramente realizou-se a decimação dos dados, esta etapa é detalhada na Seção 3.1. A escolha da taxa de decimação $\Delta$ foi realizada determinando os primeiros mínimos $r_{y^*}$ e $r_{y^{*2'}}$, Equações (\ref{eq:8}) e (\ref{eq:9}), em que o menor desses mínimos será o valor de trabalho. Através da Figura \ref{autocov} pode-se visualizar o valor de 43 amostras como o primeiro mínimo, 
sendo assim $\Delta$ = 4 foi definido como a taxa de decimação. 

\begin{figure}[htb]
	\begin{center}
		\includegraphics [scale=0.45]{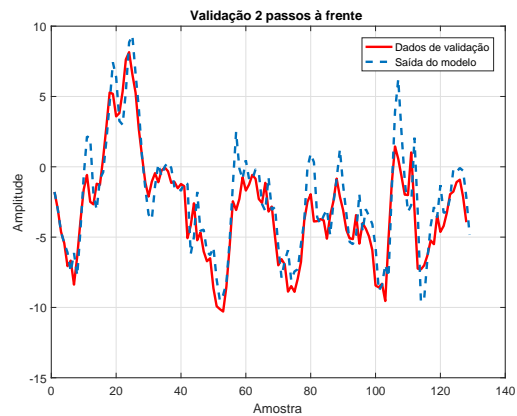}
		\caption{Saída 9 decimada, modelo NAR com predição 2 passos à frente. A linha azul (-) se refere a saída do modelo. A linha vermelha se refere aos dados de validação. O valor de RMSE entre os dois modelos foi de 0,3434.} 
		\label{Fig1}
	\end{center}
\end{figure}

\begin{figure}[H]
	\begin{center}
		\includegraphics [scale=0.5]{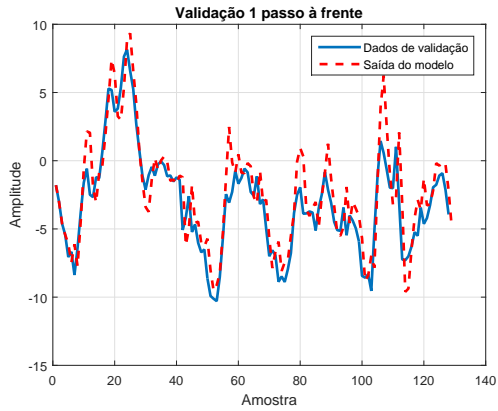}
		\caption{Saída 9 decimada, modelo NAR com predição 1 passo à frente. A linha azul se refere aos dados de validação. A linha vermelha (-) se refere a saída do modelo. O valor de RMSE entre os dois modelos foi de 0,2827.} 
		\label{Fig2}
	\end{center}
\end{figure}

O modelo obtido pelo critério de informação de Akaike, bem como os parâmetros determinados pelo estimador de mínimos quadrados (ortogonais para este caso) são mostrados na Equação (\ref{eq:3}) e as comparações entre os dados de validação para as predições 2 passos à frente e 1 passo à frente do modelo se encontram nas Figuras \ref{Fig1} e \ref{Fig2} respectivamente.

\begin{table*}[!ht]
	\caption{Taxa de Redução de Erro para os 10 Primeiros Regressores. Calculou-se a taxa de redução de erro (ERR) para
    cada termo candidato, definindo então a ordem de inserção dos mesmos no modelo. Os 4 termos com maior taxa ERR foram utilizados. }
	\begin{center}
		\label{tab:tabela1}
		\begin{tabular}{|c|c|c|c|c|c|c|c|}
			\hline
			\textbf{Termo}   & \textbf{ERR (\%)} &    \textbf{Termo}  & \textbf{ERR (\%)}    \\
			
			\hline
			$y(k-1)$ & $95,12\%$ &  $y(k-2)^3\cdot y(k-1)$& $0,060\%$  \\
			\hline
			$y(k-4)$ & $1,390\%$ & $y(k-2)$ & $0,022\%$ \\
			\hline
			$y(k-3)\cdot y(k-4)$ &  $  0,076\%$ & $y(k-3)\cdot y(k-2)^2\cdot y(k-1)$ & $0,021\%$  \\
			\hline
			$y(k-1)\cdot y(k-4)^3$ & $0,084\%$ &  $Constante$ &  $0,022\%$  \\
			\hline
			$y(k-4)\cdot y(k-1)$ &  $0,024\%$ &  $y(k-4)^4$ & $0,011\%$ \\
			\hline
			
		\end{tabular}  
	\end{center}
\end{table*}

\newpage
Pela Figura \ref{aic}, foi possível determinar que um número ótimo de acordo com o AIC foi de 4 termos, conforme sugerido na Equação (\ref{eq:3}):
\begin{equation} 
\label{eq:3}
\begin{array}{clllll} 
y(k)=1,167558558566958 \, y(k-1)    \\
- 0,228113608430163 \, y(k-4) \\ 
+ 0,007968139321360 \, y(k-4) \, y(k-3) \\
- 0,000022884707726 \, y(k-4)^3 \, y(k-1)
\end{array} 
\end{equation}

\begin{figure}[H]
	\begin{center}
		\includegraphics [scale=0.5]{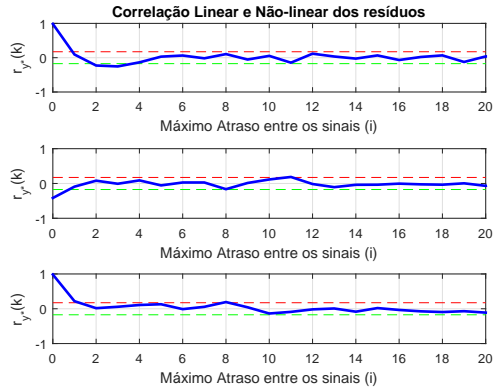}
		\caption{Autocorrelações dos resíduos. Observa-se que o modelo obtido consegue representar bem o sistema na região de operação analisada, visto que a linha em azul deve estar dentro do intervalo pontilhado na maior parte do tempo.}
		\label{residuos}
	\end{center}
\end{figure}

Ao calcular as funções de autocorrelação linear e não linear, Equações (\ref{eq:10}) $-$ (\ref{eq:12}), do vetor de resíduos $ \xi_{(xi)}$ referente ao modelo obteve-se os gráficos da Figura \ref{residuos}. O que indica que o modelo explicou tudo que era possível para este conjunto de dados. Mostrando a capacidade de predição aliada a simplicidade do modelo.

\begin{figure}[H]
\begin{center}
		\includegraphics [scale=0.3]{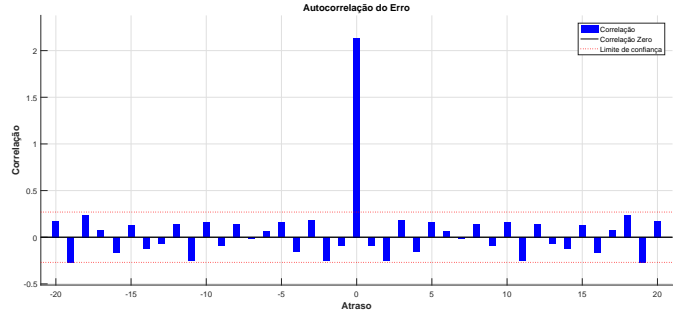}
		\caption{Autocorrelação linear do erro do modelo NAR de redes neurais. Em vermelho os limites inferior e superior do limite de confiança. Em azul a autocorrelação do erro.}
		\label{Figerro}
\end{center}
\end{figure}

\begin{figure}[htb]
\begin{center}
		\includegraphics[scale=0.3]{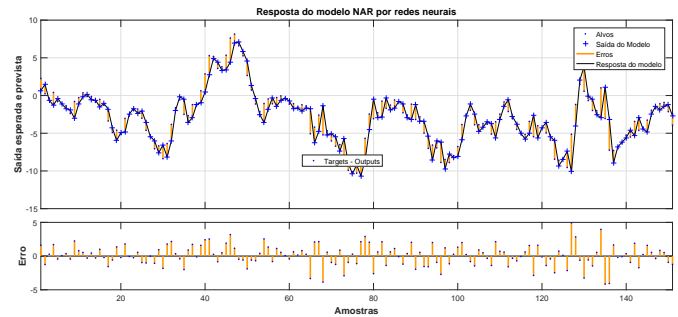}
		\caption{Saída do modelo de redes neurais. Em amarelo os erros, `+' saídas, `.' são os alvos, `-' a resposta. O valor de RMSE entre os dois modelos foi de $0,7886$. }
		\label{Figrede}
\end{center}
\end{figure}

Para a rede neural, foram observados dos testes realizados, que a rede apresenta 19 neurônios em sua camada escondida. A autocorrelação linear do erro, mostrada na Figura \ref{Figerro}, encontra-se, na maior parte do tempo, dentro dos limites de confiança. A saída do modelo para os dados de validação e teste com os dados alvo e preditos encontra-se na Figura \ref{Figrede}.

\begin{figure}[htb]
\begin{center}
		\includegraphics [scale=0.3]{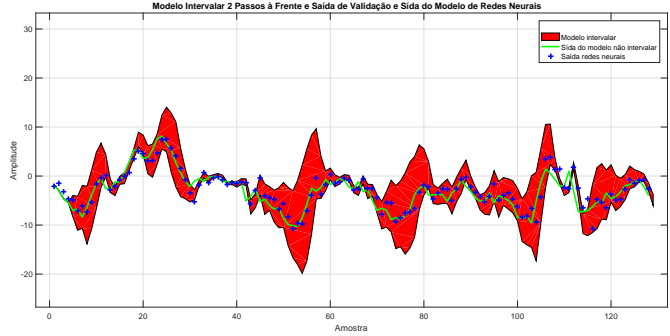}
		\caption{Comparação entre o modelo intervalar, a saída do sistema predição de 2 passos a frente e o modelo de redes neurais. Através do valor de $RMSE_{Intervalar}  = [0,1448 \,\,\,\,\,\, 0,9598]$ é possível observar que a saída intervalar representa de forma satisfatória o modelo proposto, comprovando que a variação entre os dois modelos é pequena. }
		\label{Fig3}
\end{center}
\end{figure}

De posse da estrutura do modelo identificado realizou-se a estimação dos parâmetros intervalares, para isso os dados foram inseridos em forma intervalar, ou seja, com um valor médio e um raio, como definido na Seção 3.2, na etapa de estimação de parâmetros.
A Figura \ref{Fig3} representa a saída do modelo intervalar em comparação com a saída do sistema. 
O RMSE intervalar foi obtido utilizando os dados intervalares e como mostrado na Figura \ref{Fig3} o intervalo obtido contém o valor $0,3434$ para o modelo NAR 2 passos à frente e $0,7886$ para as redes neurais calculado anteriormente para os casos não intervalares. 

Os valores dos parâmetros intervalares ${\hat{\theta}_{int}}$ se encontram a seguir:
\begin{center}
		{\small 
		${\hat{\theta}_{int}}$
		$= \left[\begin{array}{rrrr}
		  1,07616483118805, \,\,\,\,	1,25895228595358 \\
	\,\,	-0,33934032312346, \,\,\,\,\-0,11688689377745\\
		-0,	01680363479898, \,\,\,\, 0,03273991348043\\
		-0, 00011201473080, \,\,\,\, 0,00006624531590\\
		\end{array}\right]$
		}
\end{center}

Ao analisar o resultado da Equação (\ref{eq:3}) e comparando com a matriz intervalar ${\hat{\theta}_{int}}$ é possível afirmar que o intervalo contém todos os parâmetros determinado pelos mínimos quadrados e redes neurais, o que mostra a eficácia do processo.

Este processo intervalar implica na melhora da qualidade dos resultados em aplicações que a tolerância à imprecisão, aproximação e incerteza são de suma importância, como estudos aplicados a medicina, automação, sistemas aeroespaciais, biológicos, entre outros. O estudo em questão pode diminuir essa imprecisão e levar a resultados mais confiáveis. 

\newpage
\section{Conclusões}
	
No presente trabalho foi apresentada uma abordagem para considerar incertezas paramétricas no processo de identificação de sistemas. Tais incertezas são originadas no processo de aquisição de dados e consideradas durante o processo de determinação de parâmetros. A análise intervalar é relevante pois permite um conjunto infinito de parâmetros ser avaliado como um intervalo limitado. Deste modo obtêm-se uma faixa de valores para a saída do modelo. Para reduzir erros computacionais a determinação da estrutura do modelo é feita de modo clássico, isto é, sem levar-se em conta as incertezas paramétricas nos dados. 
	
Observa-se também que o valor do RMSE encontrado para o modelo NAR e redes neurais encontram-se no intervalo obtido para tal índice no modelo intervalar. Ao analisar a Figura \ref{Fig3} pode-se observar que um maior número de pontos se encontram contidos no modelo intervalar em comparação com os pontos coincidentes do modelo na Figura \ref{Fig1}. Isso se deve ao fato de que a inclusão de todos os valores reais entre os limites do intervalo permite um conjunto infinito de parâmetros ser avaliado como um intervalo limitado, com isso pode-se conseguir uma maior margem de confiabilidade na identificação de sistemas. 

Trabalhos futuros podem incluir a identificação de outro sistema bem como uma comparação deste trabalho com o desempenho utilizando o método de Monte Carlo na determinação dos parâmetros intervalares e identificação utilizando sistemas evolutivos com análise intervalar. Pretende-se investigar formas mais eficientes para reduzir os intervalos dos parâmetros.

\section*{Agradecimentos}
Os autores agradecem o apoio da UFSJ.

%
%
\bibliographystyle{splncs04}
\bibliography{Referencias}

\end{document}